\documentstyle[preprint,aps]{revtex}
\tighten

\def\beq{\begin{equation}}
\def\eeq{\end{equation}}

\def\beqn{\begin{eqnarray}}
\def\eeqn{\end{eqnarray}}
\def\nn{\nonumber\\}

\def\etal{{\it et al.}}
\def\O{{\cal O}}
\def\qq{\vec q^{\,2} \rule{0ex}{1.7ex}}
\def\Re{\,{\rm Re}\,}
\def\Im{\,{\rm Im}\,}

\begin{document}

\title{Electric polarizability of nuclei and a longitudinal sum rule}

\author{A.I. L'vov}
\address{P.N. Lebedev Physical Institute, \\
  Leninsky Prospect 53, Moscow 117924, Russia}

\maketitle

\begin{abstract}
Recently, a longitudinal sum rule for the electric polarizability of
nuclei was used to revise a relativistic correction in a dipole sum
rule for the polarizability.  This revision is shown to be wrong
because of neglecting an asymptotic contribution in the underlying
dispersion relation.  The status and correct use of the longitudinal
sum rule is clarified.

\end{abstract}

\section{Introduction}

The electric $\bar\alpha$ and magnetic $\bar\beta$ polarizabilities of
hadronic systems are currently a subject of many experimental and
theoretical studies. The polarizabilities characterize a response of
internal degrees of freedom of the system to external soft
electromagnetic fields and can be measured in reactions with real or
virtual photons \cite{mainz97}.

It was established long ago \cite{petr61} that
\beq
\label{alpha}
   \bar\alpha = \alpha_0 + \alpha_{\rm rec},
\eeq
where
\beq
\label{alpha0}
  \alpha_0 = 2 \sum_{n\ne 0}
    \frac { |\, \langle n|D_z|0\rangle \,|^2} {E_n - E_0}
\eeq
is a sum over excited states $|n\rangle$ of the system, with $\vec D$
being the operator of the electric dipole moment, and
\beq
\label{delta}
  \alpha_{\rm rec} = \frac{Z^2e^2}{3Mc^2} r_E^2
\eeq
is a relativistic recoil correction of order $\O(c^{-2})$ which is
determined by the mass $M$, the electric charge $Ze$, and the electric
radius $r_E$ of the system.  We imply in Eq.~(\ref{alpha0}) that a
disconnected contribution of the electric polarizability of the vacuum
is subtracted \cite{lvov93}, so that $\alpha_0$ is not necessarily
positive.  For the sake of clarity, we keep in formulas the speed of
the light, $c$.  We explicitly consider only the spinless case.  When
the system's spin $=1/2$, a few minor modifications are necessary in
(\ref{delta}) and in equations like (\ref{LET0}), (\ref{LET}), or
(\ref{L}) below.  They do not affect $\O(c^{-2})$ terms discussed in
this paper.  Since the term $\alpha_0$ can be written as a weighted
integral of the unretarded electric dipole cross section, we will refer
to Eq.~(\ref{alpha}) as the dipole sum rule.

Recently, Eq.~(\ref{alpha}) has been questioned.  Evaluating a
longitudinal sum rule for the electric polarizability \cite{such72} in
the case of a weakly-relativistic system (nucleus), Bernabeu
\etal\ \cite{bern98} found, apart from the terms $\alpha_0$ and
$\alpha_{\rm rec}$, an additional relativistic correction which is
driven by a non-local (i.e.\ momentum-dependent or
charge-exchange) part of the binding potential of the system and is of
the same order $\O(c^{-2})$ as the recoil term $\alpha_{\rm rec}$.

The present paper is aimed to clarify the status of the longitudinal
sum rule. We demonstrate that the sum rule used in the manner of
Ref.~\cite{bern98} is inapplicable because of a violation of the
underlying unsubtracted dispersion relation.  When treated correctly,
with an appropriate subtraction, the longitudinal sum rule leads to a
result which is free from the potential-dependent correction and is
consistent with the dipole sum rule.

\section{Dipole and longitudinal sum rules}
\label{sect2}

We begin with reminding steps leading to both the dipole and
longitudinal sum rules.  Let us introduce the Compton tensor
$T_{\mu\nu}$ which determines the amplitude of real or virtual photon
scattering, $T= \epsilon_\mu^{\prime *} T^{\mu\nu} \epsilon_\nu$.  In
the following, we need only the 00-component of $T_{\mu\nu}$ found in
the case of forward photon scattering on the system at rest,
\beq
\label{T00}
 T_{00}(\omega, q^2) = \sum_n
  \frac{ \displaystyle \Big|\, \langle n,\vec q\,|
  \int\! \rho(\vec r\,) \exp(i\vec q\cdot\vec r\,)\,d^3r \,
   |0\rangle \,\Big|^2} {E_{n,q} - E_0 - \omega-i0^+}
 + {\omega \to -\omega \choose \vec q \to -\vec q}.
\eeq
Here $q$ is the 4-momentum of the photon and $\omega=cq_0$.  Also,
$\rho(\vec r\,)$ is the charge density, $|0\rangle$ is the initial
state of the energy $E_0=Mc^2$,  and $|n,\vec q \,\rangle$ are
intermediate states of energies $E_{n,q}=\sqrt{E_n^2 + c^2\qq}$ and of
the momentum $\vec q$.  All the states are normalized as to have one
particle in the whole volume $\displaystyle \Omega=\int d^3r$. We use
the Gaussian units for electric charges, $e^2\simeq \hbar c/137$.  This
removes $1/4\pi$ in equations like (\ref{alpha0}), (\ref{delta}),
(\ref{LET0}) or (\ref{LET}) below.

The low-energy behavior of $T_{00}$ directly follows from
Eq.~(\ref{T00}). Separating the singular contribution of the unexcited
state, $n=0$, and expanding the exponent in (\ref{T00}), one has
\beq
\label{n=0}
   T_{00}(\omega, q^2) = T_{00}^{(n=0)}(\omega, q^2)
   + \qq \alpha_0 + \O(\omega^2\qq, \, \vec q^{\,4}),
\eeq
where $\alpha_0$ is given by Eq.~(\ref{alpha0}) with $\displaystyle
\vec D = \int\!  \vec r\,\rho(\vec r\,)\,d^3r$.  The
$(n=0)$-contribution is determined by the following covariant matrix
element of the electromagnetic current $j^\mu =(c\rho,\vec\jmath\,)$:
\beq
\label{j}
   \langle 0,\vec q \,|\, j^\mu(0) \, |0\rangle
  =  \frac{ceF(\tau)}
   {\sqrt{ (2E_{0,q}\Omega) (2E_0\Omega) }} ( E_{0,q}+E_0,~c\vec q\,).
\eeq
Here $F(\tau)$ is the electric form factor of the system which depends
on the momentum transfer squared $\tau = -\qq + (E_{0,q}-E_0)^2/c^2$
and is normalized as $F(0)=Z$.  Accordingly,
\beq
\label{T(n=0)}
   T_{00}^{(n=0)}(\omega, q^2)
    = e^2 F^2(\tau) \frac{(E_{0,q}+E_0)^2}{2E_{0,q}E_0}
      \, \frac{E_{0,q}-E_0}{(E_{0,q}-E_0)^2 - \omega^2}.
\eeq
When $q^2 \equiv q_0^2 - \qq =0$, Eqs.~(\ref{n=0}) and (\ref{T(n=0)})
give
\beq
\label{LET0}
   T_{00}(\omega, 0) = -\frac{Z^2e^2}{Mc^2} \Big(1-\frac{\omega^2}{3c^2}
         r_E^2 \Big) + \frac{\omega^2}{c^2} \alpha_0 + \O(\omega^4),
\eeq
where the electric radius of the system characterizes the slope of the
form factor, $F'(0)=Zr_E^2/6$.

On the other hand, using an effective Lagrangian for a polarizable
system \cite{lvov93}, one can relate $T_{00}$ with the commonly used
polarizabilities $\bar\alpha$ and $\bar\beta$ which by definition
determine a deviation of the low-energy Compton scattering amplitude
from the so-called Born term,
\beq
\label{LET-ini}
   T_{00}(\omega, q^2) = T_{00}^{\rm Born}(\omega, q^2)
   + \qq \bar\alpha + \O(\omega^2\qq, \, \vec q^{\,4}).
\eeq
The Born term describes photon scattering off a rigid (unpolarizable)
charged ``particle" of a finite size and is given by tree Feynman
diagrams with propagators and electromagnetic vertices taken in the
particle-on-shell regime. Up to the substitute $e\to eF(q^2)$, the Born
amplitude coincides with the (virtual) Compton scattering amplitude in
the scalar QED.  In the case of forward scattering,
\beq
\label{Born}
   T_{00}^{\rm Born}(\omega, q^2) = -\qq F^2(q^2) \frac{e^2}{Mc^2}\,
   \frac{4M^2c^2-q^2} {4M^2\omega^2 - (q^2)^2}.
\eeq
Therefore, Eqs.~(\ref{LET-ini}) and (\ref{Born}) give
\beq
\label{LET}
   T_{00}(\omega, 0) = -\frac{Z^2e^2}{Mc^2}
    + \frac{\omega^2}{c^2} \bar\alpha + \O(\omega^4).
\eeq
Matching Eq.~(\ref{LET}) with (\ref{LET0}), one finally obtains
\cite{petr61} the dipole sum rule (\ref{alpha}).

In essence, the dipole sum rule follows from Eq.~(\ref{T00}) which, in
turn, is valid when elementary constituents of the system have local
couplings to the electromagnetic potential and, correspondingly, a
seagull contribution $S_{00}$ to the 00-component of the Compton tensor
vanishes \cite{brow66}. The same condition of the locality is required
to suppress an exponential growth of the Compton scattering amplitude
at high complex $\omega$ and to ensure the validity of dispersion
relations.

The longitudinal sum rule just follows from matching Eq.~(\ref{LET})
with a dispersion relation for $T_{00}(\omega,0)$ which is known to be
an even analytical function of $\omega$.  The imaginary part of
$T_{00}$ can be found from (\ref{T00}).  Physically, it is related with
the longitudinal cross section $\sigma_L$ measured in
electroproduction.  Indeed, considering the amplitude of forward
virtual photon scattering $T_L=\epsilon_L^{\mu\,*} T_{\mu\nu}
\epsilon_L^\nu$ with $q^2<0$ and with the longitudinal polarization
\beq
  \epsilon_L = \frac{1}{\sqrt{-q^2 \qq}}
       (\qq;\, q_0 \vec q), \quad q\cdot\epsilon_L=0,
\eeq
and doing the gauge transformation $\tilde\epsilon_L^\mu =
\epsilon_L^\mu - \lambda q^\mu$ with an appropriate $\lambda$, one can
remove the space components of $\epsilon_L$ and arrive at the
polarization vector
\beq
  \tilde\epsilon_L = \sqrt{\frac{-q^2}{\qq}} (1;\, \vec 0).
\eeq
Accordingly,
\beq
  T_L(\omega,q^2) = \frac{-q^2}{\qq} T_{00}(\omega,q^2).
\eeq
The standard longitudinal cross section reads
\beq
  \sigma_L(\omega,q^2)=\frac{4\pi}{q_{\rm eff}} \Im T_L(\omega,q^2),
  \quad q_{\rm eff}=q_0 + \frac{q^2}{2Mc},
\eeq
so that
\beq
  \Im T_{00}(\omega, 0) = \frac{\omega^3}{4\pi c^3}
    \lim_{q^2\to 0^-} \frac{\sigma_L(\omega,q^2)}{-q^2}.
\eeq
Both $\sigma_L(\omega,q^2)$ at $q^2\le 0$ and $\Im T_{00}(\omega,0)$
are manifestly positive.

Now, {\em assuming} an unsubtracted dispersion relation for
$\omega^{-2} T_{00}$,
\beq
\label{DR}
  \omega^{-2}\Re T_{00}(\omega,0)
   = -\frac{Z^2e^2}{Mc^2\omega^2} + \frac{2}{\pi}
    {\rm P}  \int_{\omega_{\rm th}}^\infty \Im T_{00}(\omega',0)
      \frac{d\omega'}{\omega'({\omega'}^2-\omega^2)},
\eeq
one finally gets an unsubtracted longitudinal sum rule for $\bar\alpha$
which has been first suggested by Sucher and then rediscovered by
Bernabeu and Tarrach \cite{such72}:
\beqn
\label{L}
  \bar\alpha &=& \frac{2c^2}{\pi} \int_{\omega_{\rm th}}^\infty
      \Im T_{00}(\omega,0)\frac{d\omega}{\omega^3}
  = \frac{1}{2\pi^2 c} \int_{\omega_{\rm th}}^\infty
      \lim_{q^2\to 0^-} \frac{\sigma_L(\omega,q^2)}{-q^2}\,d\omega
\nn && \quad
   = \sum_{n \ne 0} \frac{2c^2}{\omega^3_n}
   \Big( 1+\frac{\omega_n}{Mc^2} \Big)
   \Big|\, \Big\langle n,\, q_z=\frac{\omega_n}{c} \Big|
  \int\! \rho(\vec r\,) \exp(i\frac{\omega_n}{c} z)\,d^3r
   \, \Big| 0 \Big\rangle \,\Big|^2,
\\ &&
   \hspace{12em} \omega_n=\frac{E^2_n-E_0^2}{2E_0}  \nonumber .
\eeqn
This sum rule should be used cautiously. Generally, the assumption that
$\omega^{-2}T_{00}$ (or $T_L(\omega,q^2)$) vanishes at $\omega=\infty$
is not valid.  Since the sum rule (\ref{L}) necessarily gives a
positive right-hand side, it is certainly violated whenever $\bar\alpha
< 0$.  Such a situation happens, for example, for polarizabilities of
pions in the pure linear $\sigma$-model (i.e.\ in the $\sigma$-model
without any heavy particles except for the $\sigma$) which are known to
be negative. To one loop, $\bar\alpha_{\pi^\pm} = 4\bar\alpha_{\pi^0} =
-e^2/(24\pi^2 m_\pi f_\pi^2) \simeq -2\cdot 10^{-4}\,{\rm fm}^3$
\cite{lvov81}.  Another instructive situation, where the sum rule
(\ref{L}) is violated, is QED, in which the amplitude $T_L$ of $\gamma
e$ scattering has a nonvanishing limit when $\omega\to\infty$
\cite{tarr78}.  At last, in the hadron world, in which one can rely on
the Regge model, the longitudinal amplitude $T_L$ of forward
scattering, as well as an amplitude with a transverse polarization, is
expected to behave as $\omega^{\alpha_R(0)}$ with the Regge-trajectory
intercept $\alpha_R(0) \simeq 1$.  Therefore,
\beq
\label{Regge}
   T_{00}(\omega, q^2) \sim \omega^{\alpha_R(0)+2}
  \quad \mbox{when $\omega\to\infty$, $q^2=$fixed},
\eeq
and a subtraction in (\ref{DR}) is badly needed.  Compared with a
high-energy behavior of transverse components of the Compton tensor
$T_{\mu\nu}$, the asymptotics (\ref{Regge}) has an additional power of
2. This perfectly agrees with a presence in (\ref{T00}) of a
time-component of the electromagnetic current $j_\mu$ which carries an
additional Lorentz factor $\sim\omega$ in its matrix elements, cf.\
Eq.~(\ref{j}).

Since the unsubtracted dispersion relation for $\omega^{-2}
T_{00}(\omega,0)$ is generally invalid, the integral in (\ref{L}), even
if convergent, does not necessarily coincide with $\bar\alpha$.  Using
in the dispersion relation a Cauchy loop of a finite size (a closed
semi-circle of radius $\omega_M$), we still can have the same
dispersion integral, though truncated at $\omega=\omega_M$. But then
the integral has to be supplemented with an asymptotic contribution
which comes from the upper semi-circle (cf.\ Ref.~\cite{petr81}).
Accordingly,
\beq
\label{L'}
   \bar\alpha =  \frac{2c^2}{\pi} \int_{\omega_{\rm th}}^{\omega_M}
     \Im T_{00}(\omega,0)\frac{d\omega}{\omega^3}
    + \bar\alpha^{(\rm as)},
\eeq
where
\beq
\label{as}
   \bar\alpha^{(\rm as)} = \frac{c^2}{\pi} \Im
   \int_{\omega=\omega_M \exp(i\phi),~0<\phi<\pi}
      T_{00}(\omega,0)\frac{d\omega}{\omega^3}.
\eeq
In particular, if the cross section $\sigma_L$ vanishes at high $\omega
\sim \omega_M$ and $T_{00}(\omega,0)$ behaves as a {\em real} function
\beq
\label{T00asym}
  T_{00}(\omega,0) \simeq  \cdots + \omega^{-2} A_{-2} +
    A_0 + \omega^2 A_2 + \omega^4 A_4 + \cdots,
\eeq
the asymptotic contribution (\ref{as}) is given by the coefficient
$A_2$,
\beq
   \bar\alpha^{(\rm as)} = c^2 A_2.
\eeq
The subtracted form (\ref{L'}) of the longitudinal sum rule should be
used instead of (\ref{L}) when $\omega^{-2}T_{00} \not\to 0$ at high
$\omega$.

We conclude this section with the statement that both the dipole and
longitudinal sum rules appear from the same quantity $T_{00}$ and they
must agree with each other.  This is explicitly checked in the next
section, in which Eq.~(\ref{L'}) is evaluated to order $\O(c^{-2})$.

\section{Longitudinal sum rule for nuclei and the asymptotic contribution}
\label{sect3}

For a weakly-relativistic system like a nucleus, the sum $I$ in
(\ref{L}) can be saturated with low-energy nonrelativistic excitations.
Introducing a generic mass $m$ of constituents and their velocity $v$,
we have $\omega_n \simeq E_n-E_0 \sim mv^2$ and $\omega_n r \sim v \ll
1$ for essential intermediate states in Eq.~(\ref{L}).  Accordingly, we
can apply a $v/c$ expansion of the relevant matrix elements,
\beqn
\label{C}
  && \Big\langle n, q_z=\frac{\omega}{c} \,\Big| \int\! \rho(\vec r\,)
   \exp(\frac{i\omega}{c}z)\,d^3r \, \Big| 0 \Big\rangle
\nn && \qquad\qquad
  = eZ \delta_{n0}
 + \frac{i\omega}{c}      \langle n|C_1(\omega)|0\rangle
 - \frac{ \omega^2}{2c^2} \langle n|C_2(\omega)|0\rangle
 - \frac{i\omega^3}{6c^3} \langle n|C_3(\omega)|0\rangle + \cdots.
\eeqn
Here $C_k(\omega) = \displaystyle \int\! (z-R_z(\omega))^k \rho(\vec
r\,) \, d^3r$ for any integer $k$, and the diagonal-in-$n$ operator
$\vec R(\omega)$ relates intermediate states $n$ at rest and at the
momentum $q_z=\omega/c$. In terms of the boost generator $\vec K$ of
the Poincare group which leaves the mass operator ${\cal M} c^2 =
\sqrt{H^2 - c^2 \vec P^2}$ of the system invariant, $\vec R(\omega)$ is
given by the equation $\displaystyle \vec R(\omega) = -\frac{c\vec K}
{\omega} {\rm arcsh} \frac{\omega} {{\cal M} c^2}
\rule[-2.5ex]{0ex}{1ex}$. This equation is a paraphrase of $|n,q_z
\rangle = \exp(-i\eta_n K_z) |n \rangle = \exp(iq_z R_z(\omega)) |n
\rangle$, where $\eta_n={\rm arcsh} (\omega/E_n)$ is a rapidity.  In
the nonrelativistic limit, $\vec R(\omega)$ becomes an ordinary
$\omega$-independent operator of the center of mass.  It is useful to
notice that $\langle n|C_1(\omega)|0\rangle = \langle n|D_z|0\rangle$
when $n\ne 0$.

Using Eq.~(\ref{C}), expanding $\omega_n$ in generic powers of $v/c$,
and keeping in Eq.~(\ref{L}) terms up to order $\O(v^2/c^2)$, we find
that $I=I_0+I_2$ where
\beq
   I_0 = 2 \sum_{n\ne 0}
    \frac { |\langle n|D_z|0\rangle|^2} {E_n - E_0},
\eeq
with the wave functions, energies, and the charge density including
$v^2/c^2$ corrections, and~%
\footnote{A similar equation of Ref.~\cite{bern98} contains also a few
terms $\sim(E_n-E_0)^2/M$ which are of order $\O(c^{-4})$ and hence can
be omitted.}
\beqn
\label{I2}
   I_2 &=& 2 \sum_{n\ne 0} \left[
        \frac{|\langle n|C_1|0\rangle|^2}{2Mc^2}
   + \frac{1}{c^2}(E_n-E_0) \left(\frac14 |\langle n|C_2|0\rangle|^2
   - \frac13 \Re[\langle 0|C_1|n\rangle\, \langle n|C_3|0\rangle]
   \right) \right]
\nn  && \quad =
     \Big\langle 0 \Big| \,\frac{C_1^2}{Mc^2}
     + \frac{1}{4c^2}  [[C_2,H],C_2]
      - \frac{1}{3c^2}  [[C_1,H],C_3] \,\Big| 0 \Big\rangle.
\eeqn
Terms of odd order in $v/c$ vanish owing to the parity conservation.
In Eq.~(\ref{I2}) and in Eqs.\ (\ref{AA}), (\ref{AAansw}) below, we
take all the quantities, including the Hamiltonian $H$, in the
nonrelativistic approximation, when all $C_k$ are $\omega$-independent
and hermitean.

In spite of an identical notation, the quantity $I_0$ is close but not
identical to $\alpha_0$, Eq.~(\ref{alpha0}).  The sum in $I_0$ involves
intermediate states of only nonrelativistic energies $E_n-E_0 \sim
mv^2$, whereas that in $\alpha_0$ includes also antiparticle states
with $E_n-E_0\sim 2mc^2$.  In the model of a relativistic particle
moving in a binding potential, the difference between $I_0$ and
$\alpha_0$ is $\O(c^{-4})$ \cite{lvov83} and hence can be neglected in
the present context.

Since the commutators in (\ref{I2}) explicitly depend on a binding
potential $V$ of the system, authors of Ref.~\cite{bern98} have
concluded that $\bar\alpha$ explicitly depends on $V$ too.  This
conclusion contradicts the dipole sum rule and technically is wrong
because Eq.~(\ref{L'}) contains an additional asymptotic contribution
$c^2 A_2$ which cancels the $V$-dependent part of $I_2$.

To find $A_2$, we have to determine an asymptotic behavior of
$T_{00}(\omega,0)$ at energies beyond those which saturate the sums for
$I_0$ and $I_2$.  Accordingly, we consider energies $\omega$ which are
high in comparison with a typical $E_n-E_0$ but still nonrelativistic
(i.e.\ $\omega =\O(mv^2)$ and $qr=\O(v/c)$). This choice allows us to
apply the $v/c$ expansion (\ref{C}). Using
\beq
   \frac{1}{E_{n,q}-E_0-\omega} + (\omega\to-\omega) \simeq
   - \frac{2}{\omega^2}(E_n-E_0) - \frac{1}{Mc^2}
\eeq
and gathering in (\ref{T00}) all terms to order $\O(c^{-4})$, we indeed
obtain a real asymptotics $T_{00}(\omega,0)\simeq A_0 + \omega^2 A_2$
with the coefficients
\beqn
\label{AA}
   A_0 &=& -\frac{2}{c^2} \sum_{n\ne 0} (E_n-E_0)
          |\langle n| C_1 |0\rangle|^2
       - \frac{Z^2e^2}{Mc^2} + \O(c^{-4}),
\nn
   A_2 &=& 2 \sum_{n\ne 0} \Bigg[
       - \frac{|\langle n|C_1|0\rangle|^2}{2Mc^4}
   + \frac{1}{c^4}(E_n-E_0)
\nn && \qquad\quad {} \times
 \left( -\frac14 |\langle n|C_2|0\rangle|^2
   + \frac13 \Re[\langle 0|C_1|n\rangle\, \langle n|C_3|0 \rangle]
   \right) \Bigg]
   + \frac{Ze}{Mc^4}\langle 0|C_2|0\rangle
\nn && \quad =
    -\frac{I_2}{c^2} + \frac{Z^2e^2}{3Mc^4}r_E^2.
\eeqn
A contribution of antiparticles to $A_0$ is absent. It is $\O(c^{-6})$
for $A_2$ and does not affect the leading term given in (\ref{AA}).
When the system's Hamiltonian $H$ and the charge density have the form
\beq
   H = \sum_{i=1}^A \Big(m_i c^2 + \frac{\vec p_i^{\,2}}{2m_i} \Big) + V,
   \quad
   \rho(\vec r\,) = \sum_{i=1}^A  e_i \delta(\vec r - \vec r_i),
\eeq
the asymptotic coefficients (\ref{AA}) are equal to
\beqn
\label{AAansw}
   A_0 &=&  -\frac{1}{c^2} \langle 0 |\, [[C_1,V],C_1] \,| 0 \rangle
   -\sum_{i=1}^A \frac{e_i^2}{m_i c^2},
\nn
   A_2 &=&  -\frac{1}{4c^4} \langle 0 |\,[[C_2,V],C_2] \,| 0 \rangle
      + \frac{1}{3c^4} \langle 0 |\,[[C_1,V],C_3] \,| 0 \rangle.
\eeqn
The leading term $A_0$ in the asymptotics of $T_{00}$ is determined by
a sum of Thomson scattering amplitudes off the constituents of the
system and by a $V$-dependent commutator which describes an enhancement
in the Thomas-Reiche-Kuhn sum rule.  This feature is very similar to
that known for the amplitude of real Compton scattering \cite{chri75}.
Meanwhile the coefficient $A_2$ is entirely determined by $V$-dependent
commutators and it vanishes when the binding potential $V$ does not
contain momentum-dependent or charge-exchange forces.

In the general case, $A_2\ne 0$ and the electric polarizability
$\bar\alpha$ found through the longitudinal sum rule (\ref{L'}) to
order $\O(c^{-2})$ reads
\beq
  \bar\alpha = I_0 + I_2 + c^2 A_2 = I_0 + \frac{Z^2e^2}{3Mc^2}r_E^2.
\eeq
It does not contain potential-dependent commutators and agrees with the
dipole sum rule. The quantities $c^2 A_2$ and $-I_2$ are determined by
the same matrix elements, and the only difference between them comes
from the state $|n=0\rangle$ which is absent in the sum (\ref{I2}).

\section{Conclusions}

We have shown that the longitudinal sum rule for $\bar\alpha$ in the
unsubtracted form (\ref{L}) used in Ref.~\cite{bern98} is not generally
valid because of a divergence of the underlying unsubtracted dispersion
relation (\ref{DR}).  In particular, it is invalid for the case of
photon scattering on nuclei in the next-to-leading order in $1/c^2$
because the binding potential $V$ does not commute with the charge
density $\rho(\vec r\,)$.  Taking properly into account an asymptotic
behavior of the longitudinal amplitude results in an additional
asymptotic contribution $c^2 A_2$, Eq.~(\ref{AAansw}), which removes
artifacts predicted in Ref.~\cite{bern98} and brings the polarizability
$\bar\alpha$ in accordance with the dipole sum rule (\ref{alpha}).

The expansion in $v/c$, which was used in Section~\ref{sect3} to
explicitly prove an agreement of the two sum rules, cannot be applied
in case when real pion production off nuclei is taken into account
since then $v/c \sim 1$.  Still, a general analysis presented in
Section~\ref{sect2} supports such an agreement, provided the subtracted
form (\ref{L'}) of the longitudinal sum rule is used.

\acknowledgments

Author appreciates useful discussions with V.A.~Petrun'kin and
A.M.~Nathan.  This work was supported in part by the Russian Foundation
for Basic Research, grant 98-02-16534.


\begin{references}

\bibitem{mainz97}
   See, e.g., Proc. Workshop on Chiral Dynamics: Theory and
   Experiment (ChPT 97), Mainz 1997, eds. A.~Bernstein, D.~Drechsel, 
   and Th.~Walcher (Springer-Verlag, 1998), talks by
   D.~Drechsel (nucl-th/9712013),
   M.A.~Moinester and V.~Steiner (hep-ex/9801008),
   B.R.~Holstein (hep-ph/9710548),
   N.~D'Hose, and others.

\bibitem{petr61}
  V.A. Petrun'kin, Sov. Phys. JETP 13 (1961) 808;
    Nucl. Phys. 55 (1964) 197; \\
  V.M. Shekhter, Sov. J. Nucl. Phys. 7 (1968) 756.
\bibitem{lvov93}
  A.I. L'vov, Int. J. Mod. Phys. A 8 (1993) 5267.

\bibitem{such72}
 J. Sucher, Phys. Rev. D 6 (1972) 1798; \\
 J. Bernabeu and R. Tarrach, Phys. Lett. 55 B (1975) 183.
\bibitem{bern98}
 J. Bernabeu, D.G. Dumm, and G. Orlandini,  Nucl. Phys. A 634 (1998) 463.

\bibitem{brow66}
 L.S. Brown, Phys. Rev. 150 (1966) 1338.

\bibitem{lvov81}
  A.I. L'vov, Sov. J. Nucl. Phys. 34 (1981) 289.
\bibitem{tarr78}
 E. Llanta and R. Tarrach, Phys. Lett. 78 B (1978) 586.

\bibitem{petr81}
 V.A. Petrun'kin, Sov. J. Part. Nucl. 12 (1981) 278; \\
 A.I. L'vov, V.A. Petrun'kin, and M. Schumacher,
   Phys. Rev. C 55 (1997) 359.

\bibitem{lvov83}
  E.I. Golbraikh, A.I. L'vov, and V.A. Petrun'kin,
  Sov. J. Nucl. Phys. 37 (1983) 868.

\bibitem{chri75}
  P. Christillin and M. Rosa-Clot, Nuovo Cim. 28 A (1975) 29.


\end{references}
\end{document}